\documentclass[
    ,final            
  ]
  {aipproc}

\layoutstyle{8x11double}

\begin{document}

\title{The asteroseismic ground-based observational counterpart of CoRoT}


\author{K. Uytterhoeven}{
  address={Laboratoire AIM, CEA/DSM-CNRS-Universit\'e Paris Diderot; CEA, IRFU, SAp, centre de Saclay, 91191, Gif-sur-Yvette, France}
 ,altaddress={INAF-Osservatorio Astronomico di Brera, Via E. Bianchi 46, 23807 Merate, Italy}
}

\author{E. Poretti}{
  address={INAF-Osservatorio Astronomico di Brera, Via E. Bianchi 46, 23807 Merate, Italy}
}

\author{P. Mathias}{
  address={UMR 6525 H. Fizeau, UNS, CNRS, OCA, Campus Valrose, 06108 Nice Cedex 2, France }
 ,altaddress={Observatoire Midi Pyrenees, Laboratoire d'Astrophysique -LATT, 57 Avenue Azereix, 65000 Tarbes, France}
}

\author{P. Amado}{
  address={Instituto de Astrof\'{\i}sica de Andaluc\'{\i}a (CSIC), Apartado 3004, 18080 Granada, Spain}
}

\author{M. Rainer}{
  address={INAF-Osservatorio Astronomico di Brera, Via E. Bianchi 46, 23807 Merate, Italy}
}

\author{S.~Mart\'{\i}n-Ruiz}{
  address={Instituto de Astrof\'{\i}sica de Andaluc\'{\i}a (CSIC), Apartado 3004, 18080 Granada, Spain}
}

\author{E.~Rodr\'{\i}guez}{
  address={Instituto de Astrof\'{\i}sica de Andaluc\'{\i}a (CSIC), Apartado 3004, 18080 Granada, Spain}
}

\author{M.~Papar\'o}{
  address={Konkoly Observatory, P.O. Box 67, 1525 Budapest, Hungary}
}

\author{K. Pollard}{
  address={Dep.\, of Physics and Astronomy, University of Canterbury, Private Bag 4800, Christchurch, New Zealand}
}

\author{C. Maceroni}{
  address={INAF-Osservatorio Astronomico di Roma, via Frascati-33, Monteporzio Catone, Italy}
}

\author{L. Balaguer-Nu\~nez}{
  address={DAM.Universitat de Barcelona, Diagonal 647, 08028, Barcelona, Spain}
 ,altaddress={Institut d'Estudis Espacials de Catalunya, c/ Gran Capit\`a 2 - 4, 08034 Barcelona, Spain}
}

\author{I. Ribas}{
  address={Institut d'Estudis Espacials de Catalunya, c/ Gran Capit\`a 2 - 4, 08034 Barcelona, Spain}
 ,altaddress={Institut de Ci\`encies de l'Espai, CSIC, Facultat de Ci\`encies, UAB, 08193 Bellaterra, Spain}
}

\author{C. Catala}{
  address={LESIA, Observatoire de Paris, CNRS, Universit\'e Paris Diderot, 5 Place Jules Janssen, 92190 Meudon, France}
}

\author{C. Neiner}{
  address={GEPI, Observatoire de Paris, CNRS, Universit\'e Paris Diderot, 5 Place Jules Janssen, 92190 Meudon, France}
}

\author{R.A. Garc\'{\i}a}{
  address={Laboratoire AIM, CEA/DSM-CNRS-Universit\'e Paris Diderot; CEA, IRFU, SAp, centre de Saclay, 91191, Gif-sur-Yvette, France}
}

\author{the CoRoT/SWG Ground-based Observations Working Group}{
 address={}
}

\begin{abstract}
We present different aspects of the ground-based observational counterpart of the CoRoT satellite mission. We give an overview of the selected asteroseismic targets, the numerous instruments and observatories involved, and the first scientific results. 
\end{abstract}

\maketitle


\section{CoRoT's simultaneous ground-based observations}
The CoRoT/SWG Ground-Based Observations Working Group is undertaking huge efforts to complement the CoRoT space data (e.g.\, \cite{Baglin}) with simultaneous ground-based data\footnote{Observations are performed at the following observatories: European Southern Observatory (ESO), La Silla and Paranal, Chile; Observatoire de Haute Provence (OHP), France; Calar Alto Astronomical Observatory (CAHA), Spain; Mount John University Observatory (MJUO), New Zealand; Observatorio Roque de los Muchachos (ORM), La Palma, Spain; Konkoly Observatory (KO), Hungary; Observatorio de Sierra Nevada (OSN), Spain; Observatorio San Pedro M\'artir (OSPM), Mexico; Observatoire du Pic du Midi (OPM), France.} to guarantee an optimal scientific outcome of the asteroseismic part of the CoRoT
mission. Activities are strongly focused on selected targets in the seismo core program, but include also the characterisation of stars in the exoplanet fields.

\subsection{The ground-based counterpart of CoRoT's seismo core programme}
\begin{figure}
\begin{tabular}{ccc}
  \includegraphics[height=.1\textheight]{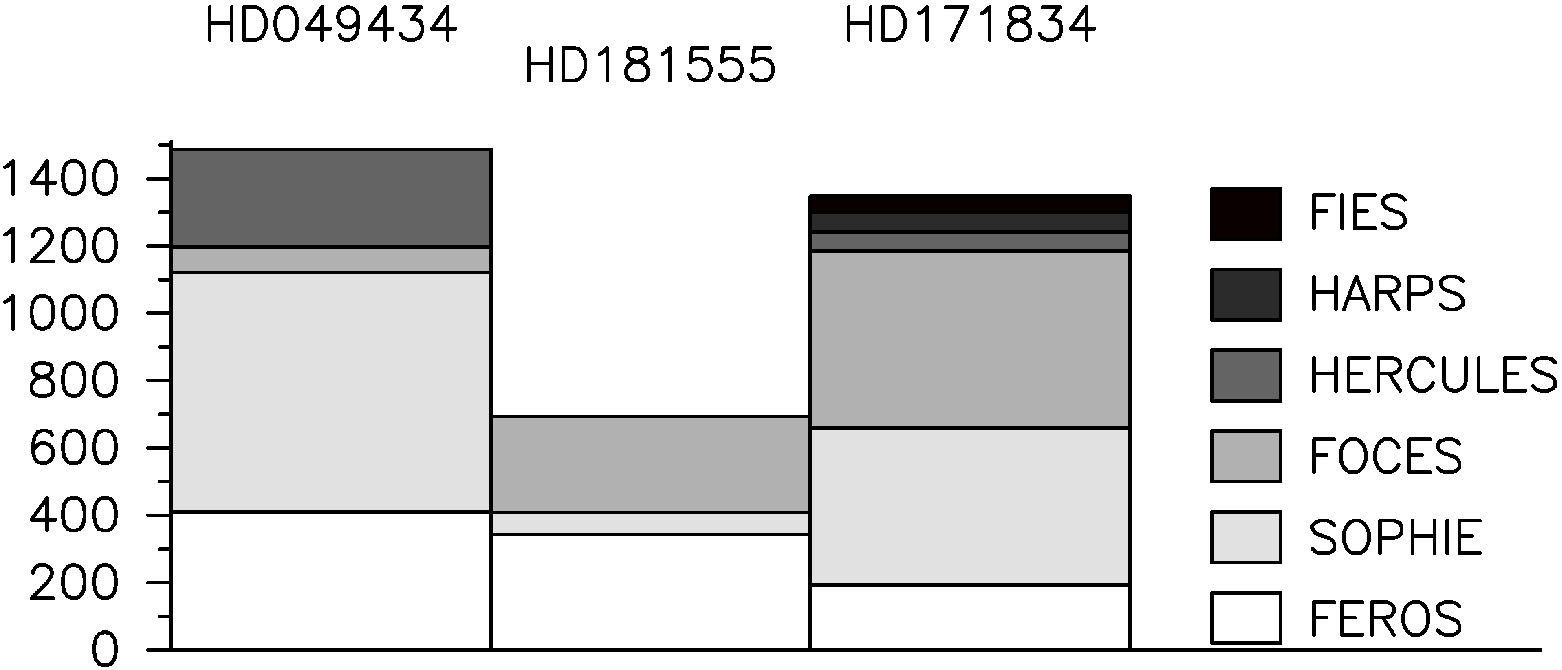} &
  \includegraphics[height=.1\textheight]{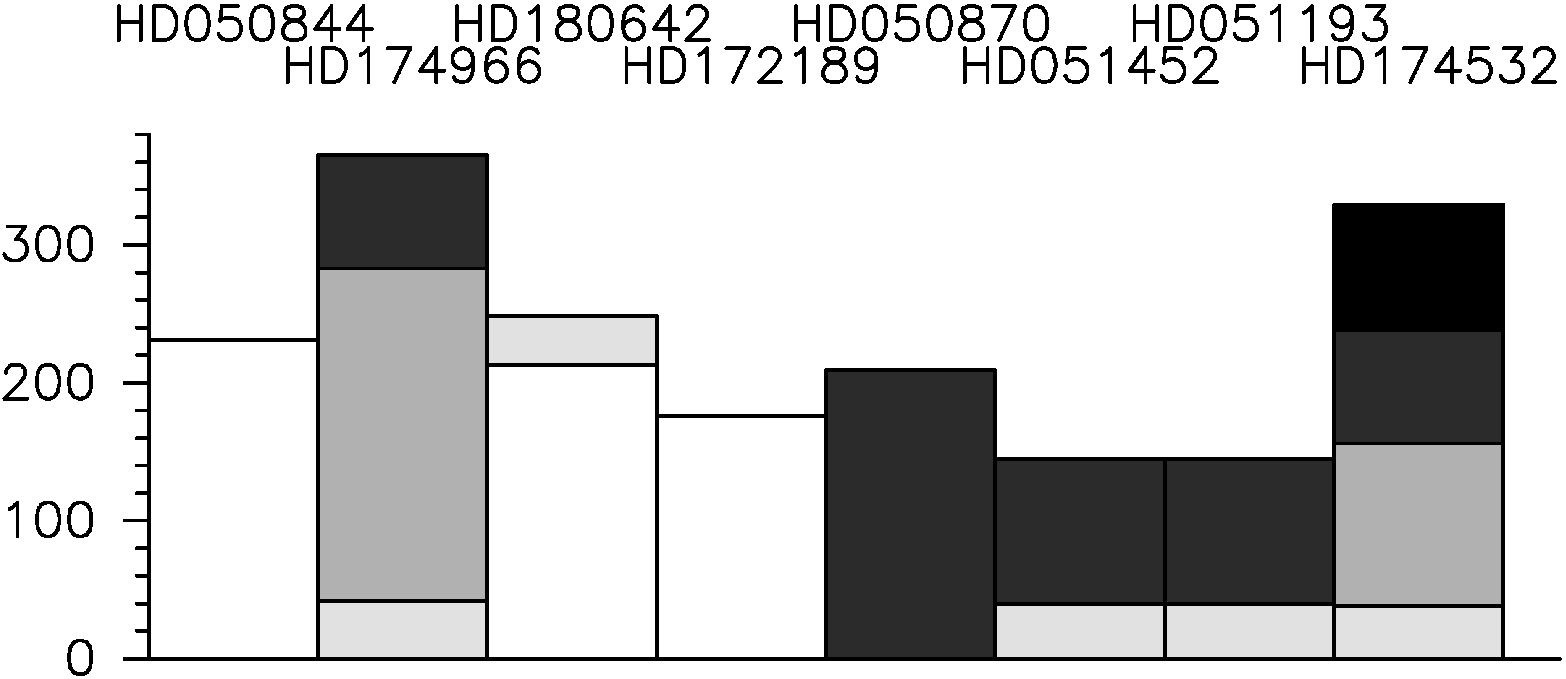} &
  \includegraphics[height=.1\textheight]{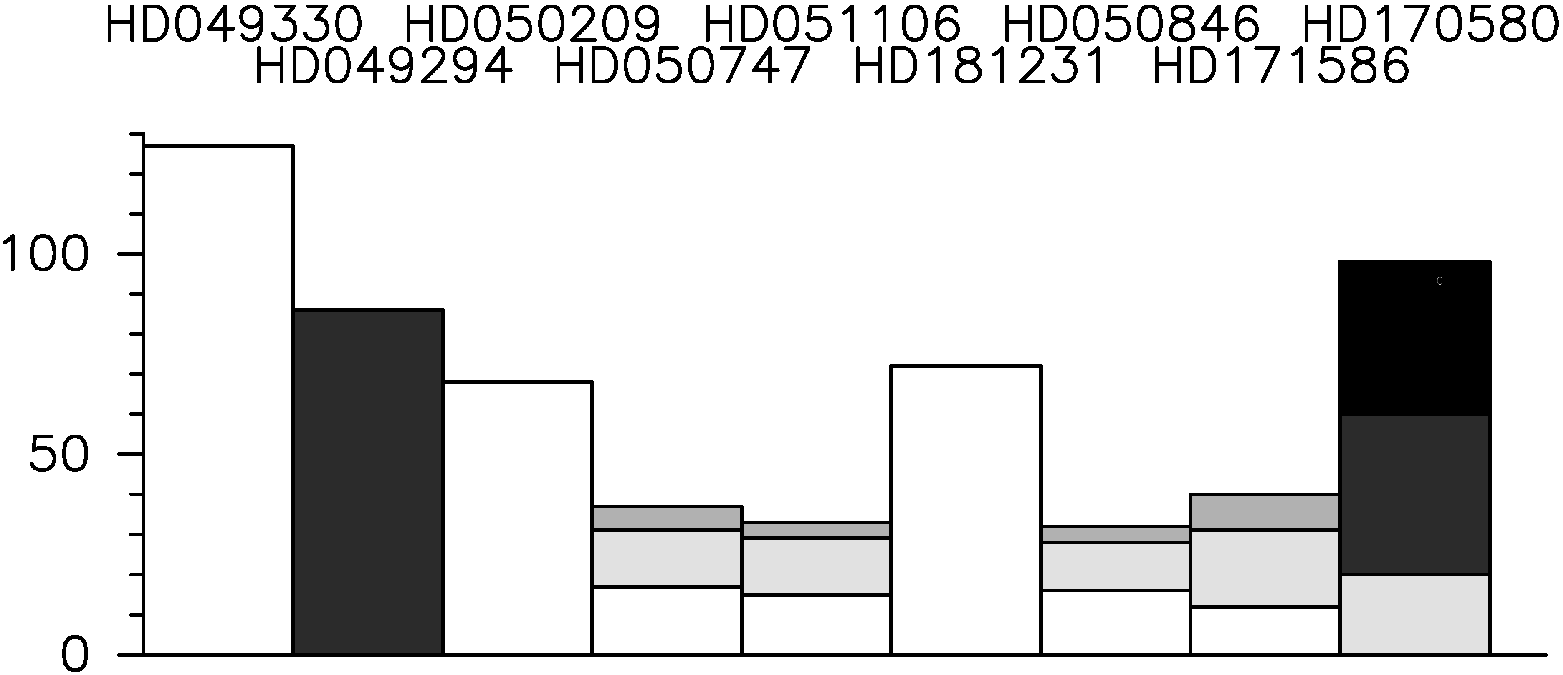} \\
\end{tabular}
  \caption{Accumulative histograms indicating the number of spectra obtained (from December 2006 to June 2009) for selected $\delta$ Sct, $\gamma$ Dor, $\beta$ Cep, Be and Ap stars with FEROS, SOPHIE, FOCES, HERCULES, HARPS and FIES. The colour code is explained in the left panel. Note the different scales on the Y-axis for the different panels. The target names are given above each column.}
\label{histograms}
\end{figure}

Several CoRoT asteroseismic targets are selected for a simultaneous follow-up from the ground. Table \ref{table1} gives an overview of the $\delta$ Sct, $\gamma$ Dor, $\beta$ Cep, Be, Ap, and  solar-like stars, eclipsing binaries (EB) and double-lined spectroscopic binaries (SB2)  that were observed between December 2006 and June 2009 (i.e. time of writing) using several high-resolution spectrographs, multi-colour photometers and/or spectropolarimeters. To ensure an unraveling of the beat frequencies multi-site campaigns have been set up, including several Large Programs. The instruments involved are listed in Table \ref{table2}. The complementary character of each project with respect to the CoRoT space data can be summarised as follows:

\begin{itemize} 
\item High-resolution spectroscopy (Panel A of  Table \ref{table2}): From time-series of high-resolution spectra valuable information on the mode parameters $(\ell,m)$ can be extracted. Figure \ref{histograms} illustrates the number of spectra obtained so far. 
\item Multi-colour photometry (Panel B of  Table \ref{table2}): Time-series of multi-colour observations provide additional information on the degree $\ell$ of the modes, that cannot be derived from the CoRoT 'white' lightcurve.
\item Spectropolarimetry (Panel C of  Table \ref{table2}): By means of spectropolarimetric measurements, signatures of magnetic fields are detected and studied.
\end{itemize}

\begin{table}
\tabcolsep=1pt 
\begin{tabular}{lrcccc|lrcccc} \hline
  \tablehead{1}{r}{b}{Target} &  \tablehead{1}{r}{b}{Type} &  \tablehead{1}{r}{b}{Field} &   \tablehead{1}{r}{b}{\# spectra} &    \tablehead{1}{r}{b}{Photom.} &    \tablehead{1}{r}{b}{TBL} & \tablehead{1}{r}{b}{Target} &  \tablehead{1}{r}{b}{Type} &  \tablehead{1}{r}{b}{Field} &   \tablehead{1}{r}{b}{\# spectra} &    \tablehead{1}{r}{b}{Photom.} &    \tablehead{1}{r}{b}{TBL}\\ \hline
  HD\,50844  &    $\delta$ Sct &    IR01  &    231  &    yes  &    - &   HD\,50170  &    solar-like &    LRa2 &    -  &    -  &    yes  \\     
  HD\,50846  &    EB &    IR01  &    32  &     -  &    - &               HD\,50870  &    $\delta$ Sct &    LRa2  &    209  &    yes  &    -  \\
  HD\,49330  &    Be  &    LRa1  &    127  &     -  &    yes  &   HD\,51452  &    Be  &    LRa2  &    145  &    -  &    yes  \\  
  HD\,49933  &    solar-like &    LRa1  &    -  &    yes  &    yes  &    HD\,51193  &    Be   &    LRa2  &    145  &    -  &    yes \\  
  HD\,49294  &    $\delta$ Sct &    LRa1  &    86  &    yes  &    - &    HD\,52265  &    solar-like &    LRa2 &    -  &    -  &    yes  \\     
  HD\,49434  &    $\gamma$ Dor &    LRa1  &    1487  &    yes  &    - &  HD170987  &    solar-like &    LRc2  &    -  &    -  &    yes  \\     
  HD\,50209  &    Be  &    LRa1  &    68  &    -  &    yes &      HD171586  &    Ap   &    LRc2  &    40  &    -  &    - \\      
  HD\,50747  &    SB2 &    LRa1  &    37  &    -  &    - &               HD171834  &    $\gamma$ Dor &    LRc2  &    1347  &     yes  &    - \\
  HD\,50773  &    Ap  &    LRa1  &    -  &    -  &    yes  &      HD172189  &    $\delta$ Sct &    LRc2  &    176  &    yes  &    - \\  
  HD\,51106  &    SB2 &    LRa1  &    33  &     -  &    - &              HD175726  &    solar-like &    LRc2  &    -  &    -  &    yes  \\     
  HD174936  &    $\delta$ Sct &    LRc1  &    -  &    yes  &    - &      HD181420  &    solar-like &    LRc2  &    - &    -  &    yes  \\      
  HD174966  &    $\delta$ Sct &    LRc1  &    365  &    yes  &    - &    HD174532   & $\delta$ Sct &    SRc2  &    329  &    -  &    - \\      
  HD180642  &    $\beta$ Cep &    LRc1  &    248  &    -  &    - &       HD\,44195  &    hybrid &    LRa3  &    -  &    yes  &    - \\         
  HD181231  &    Be  &    LRc1  &    72  &    -  &    - &         HD\,44283  &    $\delta$ Sct &    LRa3  &    -  &    yes  &    -  \\  
  HD181555  &    $\delta$ Sct &    LRc1  &    694  &    yes  &    -  &   HD170580   & $\beta$ Cep &    LRc4  &    98  &    -  &    - \\  \hline
 \end{tabular}
\caption{Overview of the targets, their variable type, and the CoRoT field. The last three columns indicate if the star is observed spectroscopically (if yes, the amount of spectra is given), photometrically and/or spectropolarimetrically.}
\label{table1}
\end{table}

\begin{table}
\tabcolsep=1pt 
\begin{tabular}{cccclc} \hline
  \tablehead{1}{c}{b}{Obs.} & \tablehead{1}{c}{b}{Instrument} & \tablehead{1}{c}{b}{Telescope} &  \tablehead{1}{c}{b}{Characteristics} &  \tablehead{1}{l}{b}{\# nights per semester/year} &   \tablehead{1}{c}{b}{{\tiny P.I.}} \\ \hline
\multicolumn{6}{l}{A. High-resolution spectrographs:} \\ \hline
 ESO & FEROS & 2.2m ESO/MPI & R$\sim$48,000 & 2006-2008: 15n per sem. & {\tiny E. Poretti} \\
 OHP & SOPHIE & 1.93m & R$\sim$70,000 & 2007-ongoing: 20n per sem. & {\tiny P. Mathias} \\
 CAHA & FOCES &2.2m & R$\sim$40,000 & 2007-ongoing: 10n per sem. & {\tiny P. Amado} \\
 ESO & HARPS &3.6m & R$\sim$80,000 & 2008-2010: 15n per sem. & {\tiny E. Poretti} \\
 MJUO & HERCULES &1.0m McLellen & R$\sim$35,000 & 2007-ongoing: 15-30n per sem. & {\tiny K. Pollard} \\
 ORM & FIES & 2.56m NOT & R$\sim$67,000 & 2009: 4n per sem.& {\tiny K. Uytterhoeven} \\
 ORM & HERMES &1.2m Mercator & R$\sim$85,000 & 2009: 6n per sem.& {\tiny K. Uytterhoeven} \\ \hline
\multicolumn{6}{l}{B. Multi-colour photometers:} \\ \hline
 OSN & Danish photometer & 0.9m & $uvby-\beta$ & 2006-ongoing: 75-100n per yr & {\tiny S. Mart\'{\i}n-Ruiz, E. Rodr\'{\i}guez} \\
 OSPM & Danish photometer& 1.5m & $uvby-\beta$ & 2006-ongoing: 18n per yr & {\tiny E. Poretti} \\
 KO & CCD & 1m RCC & BVRI & 2006-ongoing: 14-30n per sem. & {\tiny M. Papar\'o} \\ \hline 
\multicolumn{6}{l}{C. Spectropolarimeter:} \\ \hline 
 OPM & NARVAL & TBL & & 2007-ongoing: 15n per sem. & {\tiny C. Catala} \\ \hline
\multicolumn{6}{l}{D. Multi-object spectrograph and (Wide Field) CCD cameras:} \\ \hline
 ESO & FLAMES& VLT UT2 &   & 2008-ongoing: $\sim$3n per sem.  & {\tiny C. Neiner} \\ 
 ORM & WFC &2.5m INT & $uvby-\beta$ & 2007-ongoing: 2-4n per sem. & {\tiny I. Ribas, C. Maceroni, L. Balaguer, K. Uytterhoeven} \\
 KO & CCD & 1m RCC telescope & BVRI & 2008-ongoing: 14-30n per sem. & {\tiny M. Papar\'o} \\ \hline
\end{tabular}
\caption{Observatories, instruments and telescopes involved in the CoRoT ground-based campaign. The characteristics of the instruments (e.g. spectral resolution R, photometric filters) are given, as well as the amount of observing time awarded to the project. In the last column the P.I.s of the observations are listed.}
\label{table2}
\end{table}

\subsection{Characterisation of asteroseismic and binary targets in CoRoT's exoplanet fields}
The CoRoT exoplanet fields reveal a goldmine of new variable stars. However, the CoRoT lightcurves are not sufficient to characterize them. Therefore the following observing programs have been set up (Panel D of Table \ref{table2}): 

\begin{itemize} 
\item Multi-Object spectrography: Spectra are obtained for many exoplanet field stars simultaneously, which are used to identify the pulsator type and to determine fundamental stellar parameters.
\item Wide Field Camera: The complete exoplanet fields are observed in the Str\"omgren $uvby-\beta$ filters. The colours provide information on the physical parameters and reddening of the targets.
\item Multi-colour photometry: Johnson BVRI colours are observed for selected newly discovered $\delta$ Sct and RR Lyrae variables in the exoplanet fields, to characterize the stars and to provide additional information on the degree $\ell$ of the modes.
\end{itemize}

\section{First scientific results based on the observations}
For the first scientific results obtained from the ground-based observations presented here, we refer, e.g.,  to the following publications:

\begin{itemize}
\item HD\,49434: \cite{Uytterhoeven2008HD49434}
\item HD\,180642: \cite{BriquetHD180642a} 
\item HD\,50844:  \cite{Poretti2009astroph}
\item HD\,49330: \cite{Floquet}
\item HD\,50209:  \cite{Diago}
\item HD\,181231: \cite{Neiner}
\item HD\,51106 and HD\,50747:  \cite{Dolez}
\item HD\,50846 (AU Mon): \cite{Desmet}
\item Wide Field Camera data: \cite{Degroote2009}
\end{itemize}

Also, analysis results have been presented at several conferences, e.g. \cite{Uytterhoeven2008procGottingen},  \cite{Neiner08},  \cite{Gutierrez08},  \cite{Degroote}, \cite{Lefevre}, \cite{BriquetHD180642b},  \cite{Poretti}.


\begin{theacknowledgments}
This work was supported by the italian ESS project, contract ASI/INAF I/015/07/0,
WP\,03170, by the Hungarian ESA PECS project No 98022 and by the
European Helio- and Asteroseismology Network (HELAS), a major
international collaboration funded by the European Commission's Sixth
Framework Programme.  
The FEROS and HARPS data are being obtained as part of the ESO Large Programmes LP178.D-0361 and LP182.D-0356.
 KU acknowledges financial support from a \emph{European Community Marie Curie Intra-European Fellowship}, contract number MEIF-CT-2006-024476. PJA acknowledges financial
support from a Ramon y Cajal contract of the Spanish Ministry of
Education and Science.
\end{theacknowledgments}



\bibliographystyle{aipprocl} 


\IfFileExists{\jobname.bbl}{}
 {\typeout{}
  \typeout{******************************************}
  \typeout{** Please run "bibtex \jobname" to optain}
  \typeout{** the bibliography and then re-run LaTeX}
  \typeout{** twice to fix the references!}
  \typeout{******************************************}
  \typeout{}
 }

\end{document}